\documentclass[intlimits,twoside,a4paper]{article}

\usepackage{amsmath,amssymb}
\usepackage{graphicx}
\usepackage{amsmath}
\usepackage{color}

\usepackage[T2A]{fontenc}
\usepackage[cp1251]{inputenc}

\usepackage[eqsecnum]{cmpj2}

\issue{2016}{19}{2}{23001}
\doinumber{10.5488/CMP.19.23001}

\title[Hard spheres at a planar hard wall]
{Hard spheres at a planar hard wall: Simulations and density functional theory}
\author[R.L. Davidchack, B.B. Laird, R. Roth]{R.L. Davidchack\refaddr{label1}, B.B. Laird\refaddr{label2}, R. Roth\refaddr{label3}}
\addresses{
\addr{label1} Department of Mathematics, University of Leicester, Leicester, LE1 7RH, United Kingdom
\addr{label2}Department of Chemistry, University of Kansas, Lawrence, KS 66045, USA
\addr{label3}Institut f\"{u}r Theoretische Physik, Universit\"{a}t T\"{u}bingen, D-72076 T\"{u}bingen, Germany
}

\date{Received November 13, 2015, in final form December 7, 2015}
\authorcopyright{R.L. Davidchack, B.B. Laird, R. Roth, 2016}

\begin{document}

\maketitle

\begin{abstract}
Hard spheres are a central and important model reference system for both
homogeneous and inhomogeneous fluid systems. In this paper we present
new high-precision molecular-dynamics computer simulations for a hard
sphere fluid at a planar hard wall. For this system we present
benchmark data for the density profile $\rho(z)$ at various bulk
densities, the wall surface free energy $\gamma$, the excess
adsorption $\Gamma$, and the excess volume $v_\textrm{ex}$, which is closely
related to $\Gamma$. We compare all benchmark quantities with
predictions from state-of-the-art classical density functional theory
calculations within the framework of fundamental measure theory. While
we find overall good agreement between computer simulations and theory,
 significant deviations appear at sufficiently high bulk densities.
\keywords    inhomogeneous fluids, solid-liquid interfaces,  surface thermodynamics, classical density functional theory, molecular-dynamics simulation
\pacs 05.20.Jj, 05.70.Np, 61.20.Gy, 61.20.Ja, 68.08.De, 68.35.Md
\end{abstract}

\maketitle

\section{Introduction}

Hard spheres are a central model in statistical physics and act as
a reference system for simple and complex fluids. The bulk phase
behavior of the hard-sphere system is simple because the system is
athermal~--- there is no energy scale to be compared to the
thermal energy of $k_\text{B} T$, where $k_\text{B}$ is the
Boltzmann constant and $T$ is the absolute temperature. Therefore,
the phase behavior is controlled solely by the particle density
$\rho=N/V$~--- or equivalently by the packing fraction
$\eta=\rho^*\pi/6$, where the reduced density is defined as
$\rho^* = \rho \sigma^3$, with $\sigma$ being the hard-sphere
diameter.  For packing fractions $\eta < 0.492$, the equilibrium
thermodynamic phase is fluid, while above this value of $\eta$,
hard spheres are capable of forming a crystal.

In general,  the structure of simple liquids is dominated by
packing effects generated by the steep and short-ranged repulsive
  part of the interatomic interaction, which can be very well
approximated by a hard-sphere interaction. The usefulness of the
  hard-sphere system as a reference is not limited to bulk systems,
  but also extends to inhomogeneous fluids. For example, the
  system studied here, i.e., a hard-sphere fluid confined by planar hard
walls, is one of the simplest  model systems for a fluid-solid interface. 
The wall is modelled by the external potential
\begin{equation}
V_\text{ext}({\bf r}) = V_\text{ext}(z)=\left\{
\begin{array}{ll}
\infty, & \qquad z < {\sigma}/{2}, \\
0, & \qquad \text{otherwise},
\end{array}
\right.
\end{equation}
where $z$ is the distance normal to the wall. Close to the wall,
the fluid develops an inhomogeneous structure, which can be
described through the ensemble averaged density profile $\rho({\bf
r})$. At sufficiently low bulk densities, when spontaneous
symmetry breaking due to freezing can be ruled out, the
equilibrium density profile possesses the same spatial symmetry as
the external potential, so that we can assume that the density
profiles depend only on $z$, that is, $\rho({\bf r})=\rho(z)$.
From the density profile as a function of packing fraction, one
can determine a number of thermodynamic interfacial properties,
  including the excess adsorption $\Gamma$, the excess volume
  $v_\textrm{ex} = \rho \Gamma$ and the wall surface free energy~$\gamma$.

Recently,  high precision results for the excess adsorption
$\Gamma$ and the wall surface free energy
$\gamma$ at various bulk densities have been obtained using molecular-dynamics simulation
\cite{Laird07,Laird10}. Due to their high precision, these data are well suited as benchmarks to enable the testing of theoretical predictions.  In this paper we present the results for the density profile $\rho(z)$, from these high precision simulations,  together
with comparisons with state-of-the-art classical density
  functional theories (DFT) for hard-sphere fluids~--- namely DFT
  formulations based on fundamental measure theory (FMT) \cite{Rosenfeld89,Roth10}.

\section{Methods}
\label{sec:2}

\subsection{Simulation}

Density profiles of the hard-sphere fluid at a planar hard wall
were determined in the molecular-dynamics (MD) simulation using
the algorithm of Rapaport \cite{RapaportBook}. The walls were
placed normal to the $z$-axis at a distance of about $65\sigma$
apart. The $x$-$y$ cross section was approximately square with a
side length of about $50\sigma$.  Periodic boundary conditions
were employed in the $x$ and $y$ directions.  To measure the
density profiles, $\rho(z)$, the system was divided along the
$z$-axis into bins of width $0.02\sigma$. The size of the
simulation box was the same in all simulations while the number of
spheres varied from about 8\,000 for the lowest bulk  reduced
density of about 0.052 to about 150\,000 for the highest reduced
density of about 0.938. Systems at 17 different densities were
simulated.  At each density, we performed 50 independent runs
starting from well equilibrated initial states. The data for the
density profiles were averaged over the runs and over the two
walls. The 95\% confidence intervals were estimated from the
scatter in the data for independent runs.  We also simulated
systems of smaller size at a range of densities and determined
that the size of the systematic error due to the finite system
size was much smaller than the statistical errors. In this
manuscript, we include only a few examples of the density
profiles.  The complete tabulated density profiles can be found in
the Supplementary Material \cite{supplement}.  MD results for the
excess volume, $v_\textrm{ex}$, were obtained from the density
profiles using the procedure described in the Supplement to
reference~\cite{Laird12}.

Near the freezing density, the hard-sphere fluid at a hard wall is
known to exhibit a pre-freezing transition
\cite{Courtemanche92,Dijkstra04}; however, the transition is known
to have a significant nucleation barrier due to line tension
\cite{Auer03}. We have examined the structure of our hard-sphere
fluid at and near the wall surface and find no evidence of
crystalline ordering that would indicate that prefreezing has
occured, even at the highest packing fractions; therefore, we are
confident that we are examining a fully fluid-wall interface.

\subsection{Density functional theory}

In density functional theory (DFT), there exists a functional
$\Omega[\rho]$ of the density distribution $\rho({\bf r})$ of the
form \cite{Evans79}
\begin{equation} \label{eq:omega}
\Omega[\rho]={\cal F}[\rho]+\int \rd^3 r~\rho({\bf r})[V_\text{ext}({\bf r})-\mu],
\end{equation}
where ${\cal F}[\rho]$ is the functional of the intrinsic
Helmholtz free energy, $V_\text{ext}({\bf r})$ is the external and
$\mu$ the chemical potential. It can be shown that the functional
$\Omega[\rho]$ is minimized by the equilibrium density profile
$\rho_0({\bf r})$ and that it reduces to $\Omega$~--- the grand
potential of the system in equilibrium, i.e.,
$\Omega=\Omega[\rho_0]$ \cite{Evans79}. These properties can be
employed in order to obtain the inhomogeneous structure of a fluid
in an external potential within the same framework as
thermodynamic quantities. From the variational principle of DFT we
obtain the equilibrium density profile:
\begin{equation}
\left.\frac{\delta\Omega[\rho]}{\delta \rho({\bf r})}
\right|_{\rho({\bf r})=\rho_0({\bf r})} = 0.
\end{equation}

From the density profile $\rho_0({\bf r})$, in general, or $\rho_0(z)$
in the present study, one can directly compute the excess adsorption
$\Gamma$ via
\begin{equation} \label{eq:adsorp}
\Gamma=\frac{1}{A}\int_V \rd^3 r~\left[\rho_0({\bf r})-\rho\right]=\int_L \rd z~[\rho_0(z)-\rho],
\end{equation}
where $\rho$ is the bulk density, and the wall surface free energy
$\gamma$ via
\begin{equation} \label{eq:tension}
\gamma=\frac{1}{A}\left\{\Omega[\rho_0({\bf r})]+p V\right\}= \Omega[\rho_0(z)]+p L.
\end{equation}
Here, $A$ is the area of the wall, which is assumed to be infinite
and $L=V/A$ is the extension of the system in $z$-direction. In
order to make $\Gamma$ and $\gamma$ well-defined quantities, it is
necessary to define the volume $V$ in equations~(\ref{eq:adsorp})
and (\ref{eq:tension}), i.e., one must define the dividing
interface at which the system and the wall are separated
\cite{Bryk03}. Here, we use the location of the actual hard wall
as a dividing interface. As a consequence, the excess adsorption
$\Gamma$ for a hard-sphere fluid turns out to be negative and the
surface free energy $\gamma$ positive. Because the wall surface free
energy $\gamma$ and the excess adsorption $\Gamma$ are related by
Gibbs's adsorption theorem
\begin{equation} \label{eq:gibbs}
\Gamma = - \left(\frac{\partial \gamma}{\partial \mu}\right)_{T,V}
\end{equation}
the definition of $V$ or $L$ must be the same in both
equations~(\ref{eq:adsorp}) and (\ref{eq:tension}).

Another useful structural quantity is the excess interfacial volume, $v_\textrm{ex}$, defined as
\begin{equation} \label{eq:exv}
v_\textrm{ex} =\int_L \rd z~\left[1 - \frac{\rho_0(z)}{\rho}\right].
\end{equation}
This quantity, which is related to the interfacial adsorption by
\begin{equation}
\Gamma = -\rho v_\textrm{ex}
\end{equation}
is useful because it provides a convenient route to the
determination of $\gamma$ directly from the density profile
through the relation
\begin{equation}\label{eq:dgdp}
v_\textrm{ex} = \left (\frac{\partial \gamma}{\partial p}\right )_{T,N}
\end{equation}
as was illustrated in reference~\cite{Laird10}.

In order to use DFT, one must specify the functional ${\cal
F}[\rho]$ of the intrinsic Helmholtz free energy in
equation~(\ref{eq:omega}). The intrinsic Helmholtz free energy can
be split into
\begin{equation}
{\cal F}[\rho]={\cal F}_\text{id}[\rho]+{\cal F}_\text{ex}[\rho]
\end{equation}
with an exactly know ideal gas contribution
\begin{equation} \label{eq:ig}
{\cal F}_\text{id}[\rho]=\beta^{-1} \int \rd^3 r~\rho({\bf r})\left[ \ln \lambda^3
\rho({\bf r})-1\right]
\end{equation}
and an excess (over the ideal gas) contribution ${\cal F}_{ex}[\rho]$,
which contains all the information about inter-particle
interaction. In equation~(\ref{eq:ig}) $\beta=1/(k_\text{B} T)$, and $\lambda$ is
the thermal wavelength.

For the system of interest, a hard sphere fluid, fundamental
measure theory (FMT)\cite{Rosenfeld89,Roth10} provides an accurate
approach for the excess free energy functional. Within FMT, the
excess free energy is written as follows:
\begin{equation}
{\cal F}_\text{ex}[\rho]=\beta^{-1} \int \rd^3 r~\Phi(\{n_\alpha\}),
\end{equation}
where $\Phi$, the excess free energy density, is a function of weighted
densities $n_\alpha({\bf r})$ \cite{Rosenfeld89,Roth10}. The details
of FMT, including the definition of the weighted densities can be
found in a recent review \cite{Roth10}. Here, we employ three different
versions of FMT: (i) the original Rosenfeld functional
\cite{Rosenfeld89}, (ii) the White-Bear version of FMT
\cite{Roth02,Yu02}, and (iii) the White-Bear version of FMT Mark II
\cite{HansenGoos06b}. 
The main difference between these three versions of FMT are the equations of state underlying the functionals:
the Percus-Yevick (PY) compressibility equation for the Rosenfeld functional, the Manssori-Carnahan-Starling-Leland (MCSL) \cite{Mansoori71}
pressure for the White Bear version and a recently proposed, somewhat more consistent generalization \cite{HansenGoos06a}, 
of the Carnahan-Starling (CS) pressure for the White-Bear Mark II functional.

\section{Results}
\label{sec:3}

\subsection{Density profiles}

We begin by presenting density profiles of a hard-sphere fluid in
contact with a planar hard wall at selected values of the bulk density.
The density profiles from the MD simulations act as benchmark data
for a comparison with the results from different versions of FMT.

It is interesting to note that, for a fluid in contact with a planar
hard wall, the density closest to the wall, the so-called contact
density $\rho_\text{c}$, is fixed by the contact theorem
\begin{equation} \label{eq:wall}
\rho_\text{c} = \rho(z=\sigma^+/2)=\beta p,
\end{equation}
where  $p$ is the bulk pressure. Equation~(\ref{eq:wall}) is
satisfied by the results of FMT \cite{Roth10}.

It is well known that the CS pressure, which underlies the
one-component White-Bear and White Bear Mark II versions of FMT,
is more accurate, as measured by comparison to computer
simulations, than the PY compressibility pressure, which underlies
the original Rosenfeld functional. At low bulk density, the
difference between the CS and PY equation of state is small, but
at sufficiently high bulk fluid densities, the PY pressure
significantly overestimates the pressure of a hard-sphere fluid
relative to the
 computer simulation results. Hence,  it is to be expected that very
close to the wall, where $\rho(z)$ is strongly influenced by the
contact theorem [equation~(\ref{eq:wall})], the density profiles obtained from
the White-Bear versions of FMT will be closer to the simulation results
than those calculated using the Rosenfeld functional.

\begin{figure}[!b]
\centerline{
\includegraphics[width=0.55\textwidth]{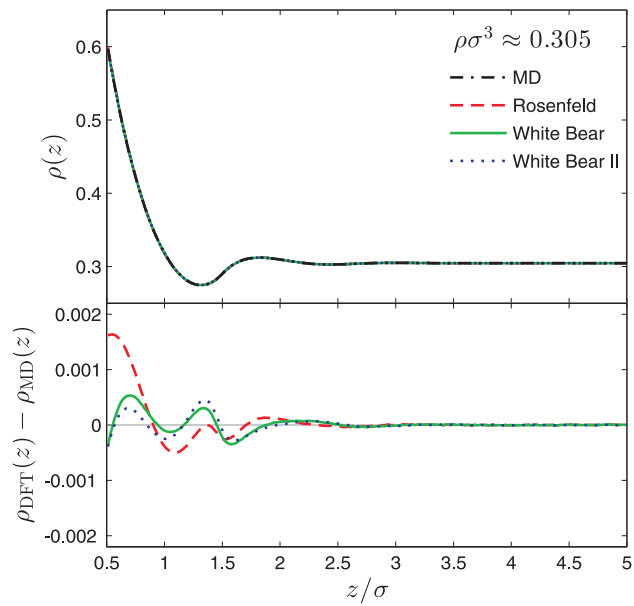}
}
\caption{\label{fig:profile1} (Color online) {Top panel}: Equilibrium density profiles $\rho(z)$ of
  a hard-sphere fluid of reduced density $\rho^* \approx 0.305$ ($\eta \approx 0.159$) at a planar hard wall.  The black line denotes data from MD simulation, while the red, green and blue lines correspond
  to various versions of FMT.  {Bottom panel}: the deviation of the FMT densities from the MD
    results. The statistical errors in the MD data are smaller than the
  width of the lines.}
\end{figure}

\begin{figure}[!t]
\centerline{
\includegraphics[width=0.545\textwidth]{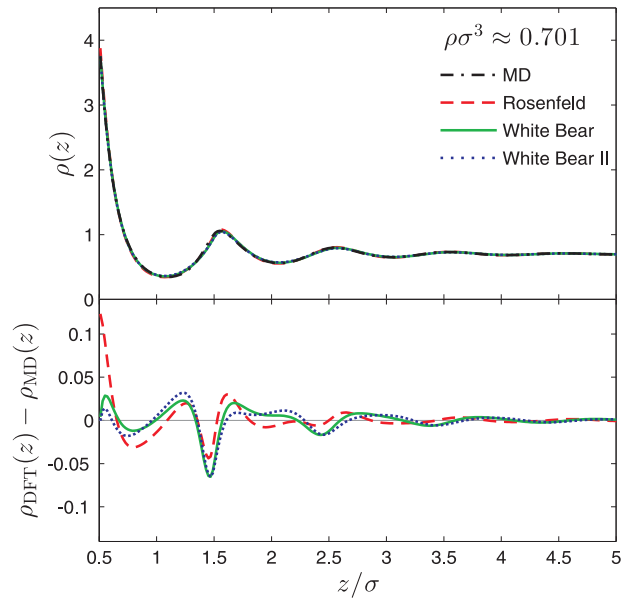}
}
\vspace{-3.5mm}
\caption{\label{fig:profile2} (Color online) Same as figure~\ref{fig:profile1} for $\rho^*\approx 0.701$ ($\eta\approx 0.367$).}
\end{figure}

\begin{figure}[!b]
\centerline{
\includegraphics[width=0.545\textwidth]{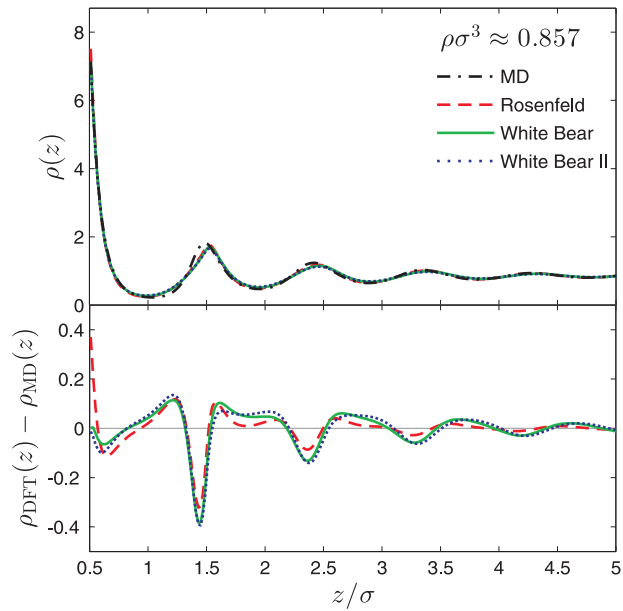}
}
\vspace{-3.5mm}
\caption{\label{fig:profile3} (Color online) Same as figure~\ref{fig:profile1} for $\rho^*\approx 0.857$ ($\eta\approx 0.449$).}
\end{figure}

\begin{figure}[!t]
\centerline{
\includegraphics[width=0.55\textwidth]{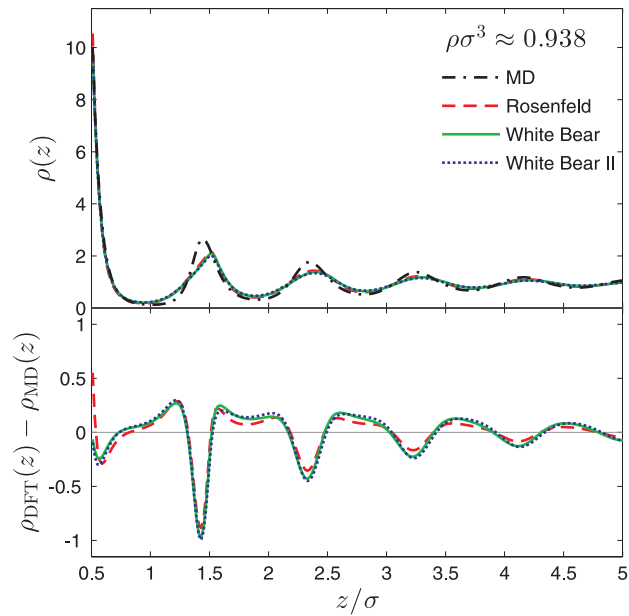}
}
\vspace{-3mm}
\caption{\label{fig:profile4} (Color online) Same as figure~\ref{fig:profile1} for $\rho^*\approx 0.938$ ($\eta\approx 0.492$).}
\end{figure}

In figures~\ref{fig:profile1}--\ref{fig:profile4}, we show density
profiles at four different values of the bulk density. In
figure~\ref{fig:profile1}, the bulk density $\rho^* \approx
0.305$,  corresponding to a bulk packing fraction of $\eta \approx
0.159$, is relatively low. All versions of FMT predict the density
profiles that agree very well over the whole range of $z$ with
that obtained from computer simulations. At the wall, as expected,
the White Bear versions of FMT are more accurate than the
Rosenfeld functional, but the difference is very small, as can be
estimated from the difference between the PY and the CS equations
of state (about 0.3\%). The packing effects close to the wall are
small and decay fast.

At a somewhat larger value of the bulk density, $\rho^* \approx
0.701$  ($\eta\approx 0.367$) (shown in figure~\ref{fig:profile2}),
the agreement between the profiles from
FMT and those from computer simulations is very good in all
details. At this intermediate bulk density, the PY and CS pressure
differ by about 3\%, which is reflected by a small
deviation between the density profile
obtained from the Rosenfeld functional and the profiles from the White
Bear versions of FMT--- especially close to the wall. This deviation, however, is sufficiently small
that it can only be seen in the bottom panel of
figure~\ref{fig:profile2}. Packing effects at this bulk density are
significantly more pronounced than for the bulk density used for
figure~\ref{fig:profile1}~--- note the different scales for the
$y$-axes in the bottom panel.

As the bulk density is further increased to $\rho^*\approx 0.857$
($\eta\approx 0.449$) (figure~\ref{fig:profile3}), and
$\rho^*\approx 0.938$ ($\eta\approx 0.492$)
(figure~\ref{fig:profile4}), respectively, the agreement between
FMT  and computer simulations  is still good very close to the
wall, but clearly less satisfying for the oscillatory
structure~--- we observe that the FMT significantly underestimates
 the height of the density peaks about one particle
diameter away from the wall. Especially at $\rho^*\approx
0.938$, a density close to bulk freezing, one can see that the height
of the second density peak is also underestimated by FMT.

Since all versions of FMT seem to have a problem with the second
peak in the density profile at high fluid density, it seems likely
that it is not due to the precise form of the excess free energy
density $\Phi$ employed, but rather the structure and range of the
weight functions that can only approximate the complicated
integrals of the virial expansion of the free energy
\cite{Leithall11,Korden12,Marechal14}.

\subsection{Surface free energy $\gamma$}

From the equilibrium density profile it is straightforward within
DFT to compute the wall surface free energy from
equation~(\ref{eq:tension}). Because various versions of FMT result
in slightly different density profiles, as shown in
figures~\ref{fig:profile1}--\ref{fig:profile4}, the wall surface
free energy obtained from different versions of FMT are expected
to differ by a small amount.

The MD simulation results were obtained, as in
reference~\cite{Laird10}, using equation~(\ref{eq:dgdp}) and
integrating the excess volume, $v_\textrm{ex}$, with respect to
pressure. In that work, the numerical error in the integration was
minimized by subtracting from the integrand the expression for the
excess volume from Scaled Particle Theory (SPT) and then adding
back, after integration, the exact expression for the SPT. For the
data used in reference~\cite{Laird10} this was sufficient to
ensure that the numerical integration error was smaller than the
reported statistical error. However, this is not sufficient for
the high precision simulations reported here. To calculate
$\gamma$ from the current MD results, instead of subtracting the
SPT expression from the integrand, we subtract the corresponding
expression from a recent high-accuracy parameterization of
$\gamma$, presented in reference~\cite{Davidchack15}. This results
in a numerical error that is smaller than the reported statistical
error. A table of numerical values of $\gamma$ is presented in the
Supplementary Information \cite{supplement}.

\begin{figure}[!t]
\centerline{
\includegraphics[width=0.5\textwidth]{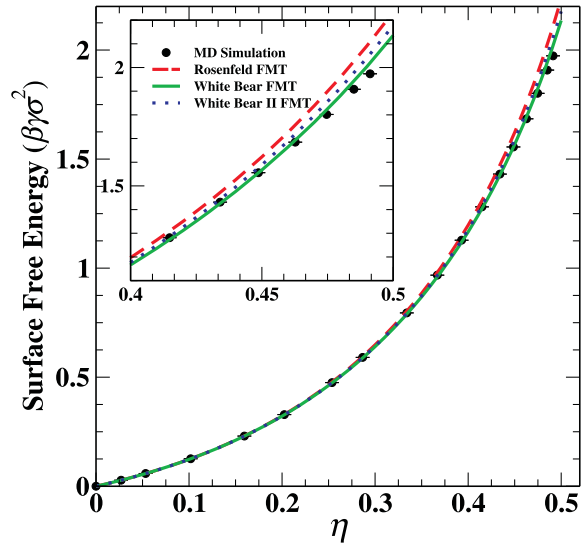}
}
\caption{\label{fig:tension} (Color online) The wall surface free energy $\gamma$ of a
  hard-sphere fluid at a planar hard wall as a function of the packing
  fraction $\eta$. The overall agreement between different
  versions of FMT (lines) and MD simulations (symbols) is
  excellent. For $\eta\geqslant 0.4$ there are small deviations between the
  DFT and MD results. }
\end{figure}

In figure~\ref{fig:tension}, we show a comparison of the wall
surface free energy as a function of the packing fraction $\eta$
from different versions of FMT (lines) and from MD simulations
(symbols). As expected, the agreement among different versions of
FMT and the MD simulations is excellent at small and intermediate
values of the packing fraction and remains good in the whole fluid
density range.

At higher values of the packing fraction, $\eta\geqslant 0.3$, one
can observe that the FMT results slightly overestimate the wall
surface free energy compared to the MD results
\cite{Davidchack15}. In the inset, we highlight the region of high
values of $\eta$, from which one can clearly see  that the wall
surface free energy obtained by the original Rosenfeld functional
(red line) overestimates the values from the simulations the most,
while the prediction from the White-Bear version of FMT (green
line) is closest to the MD data.

Recently, a new parametrization of the simulation data for
$\gamma$ was given by us \cite{Davidchack15}, that is accurate in
the whole range of fluid densities. Especially at higher
densities, an additional term with a high power in $\eta$ is
required in the parametrization in order to account for all the
simulation data. Such a term is not reflected in the excess free
energy density $\Phi$ of FMT.

\subsection{Adsorption $\Gamma$ and excess volume $v_\textrm{ex}$}

As discussed in section~\ref{sec:2}, the excess adsorption $\Gamma$ can be calculated
via two different routes: (i) by integrating the density profile,
equation~(\ref{eq:adsorp}), or (ii) using the adsorption
theorem, equation~(\ref{eq:gibbs}). We have confirmed that our DFT and MD results
are consistent, in that they yield the same values of $\Gamma$ via both routes. The MD data for $\Gamma$ can be found in the Supplemental Information \cite{supplement}.

\begin{figure}[!t]
\centerline{
\includegraphics[width=0.5\textwidth]{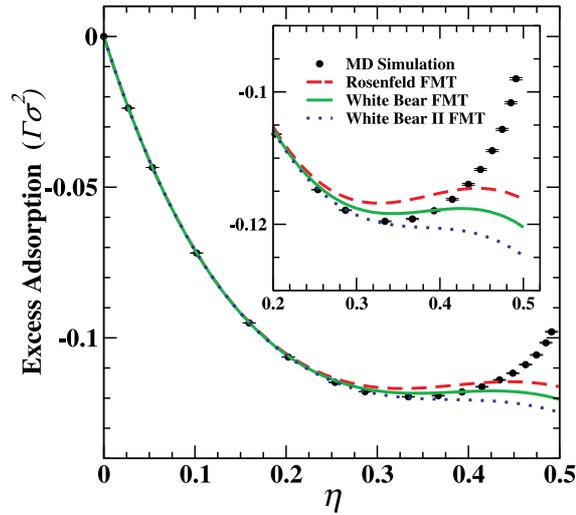}
}
\caption{\label{fig:adsorption} (Color online) The excess adsorption $\Gamma$ for a
 hard-sphere fluid at a planar hard wall as a function of the packing
  fraction $\eta$. The symbols denote data from MD simulation, while
  the lines correspond to different versions of FMT.}
\end{figure}

\begin{figure}[!b]
\centerline{
\includegraphics[width=0.49\textwidth]{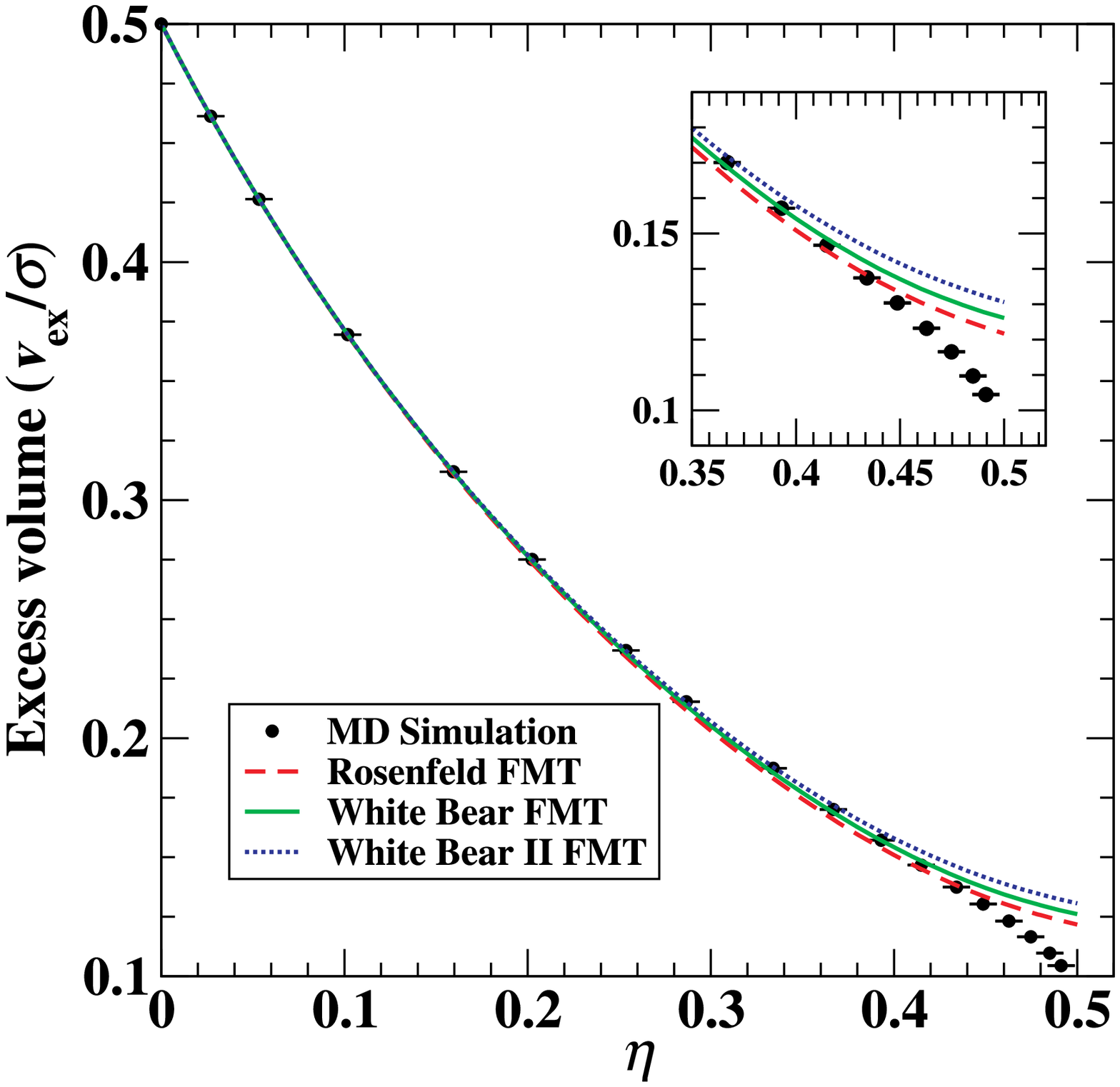}
}
\caption{\label{fig:vex} (Color online) The excess volume  $v_\textrm{ex}$ for a
 hard-sphere fluid at a planar hard wall as a function of the packing
  fraction $\eta$. The symbols denote data from MD simulation, while
  the lines correspond to different versions of FMT. }
\end{figure}

In figure~\ref{fig:adsorption}, we show the excess adsorption as a
function of the packing fraction $\eta$. At low values of $\eta$,
the DFT results are indistinguishable from each other and agree
very well with the MD data. In contrast to the wall surface free
energy $\gamma$, shown in figure~\ref{fig:tension}, FMT results
start to deviate from the MD data already at a moderate value of
$\eta\geqslant 0.25$. For $\eta\geqslant 0.3$, the results from
different versions of FMT begin to deviate from each other and for
$\eta\geqslant 0.4$, there is a significant difference between the
simulation results and those from FMT. While the excess adsorption
increases towards less negative values for $\eta\geqslant 0.4$ in
the simulation results, the FMT results display a local minimum
and then decrease further for $\eta\to 0.5$. It should be noted
that the presence of the local minimum is a consequence of the
current choice of volume definition. If one uses another commonly
used volume definition in which the position of the wall is
coincident with the center of a hard sphere in contact with the
wall then the local minimum disappears; however, the
underestimation by the DFT of $\Gamma$ at high packing fractions
remains.

The excess volume, $v_\textrm{ex}$ as a function of $\eta$ is
shown in figure~\ref{fig:vex}~--- see the Supplemental
Information\cite{supplement} for numerical values of
$v_\textrm{ex}(\eta)$ obtained from MD simulation. As is the case for $\Gamma$, there are
significant deviations in the FMT estimates for $v_\textrm{ex}$
from the MD results at high packing fraction, $\eta > 0.4$. At
these high packing fractions, the values of $v_\textrm{ex}$ are
overestimated by all versions of FMT examined. The excess volume
has a direct relationship to the surface free energy through
equation~(\ref{eq:dgdp}) and it is through the integration of this
equation that the MD simulation values for $\gamma$ shown in
figure~\ref{fig:tension} are calculated \cite{Laird10}. Because
$v_\textrm{ex}$ is directly calculated from the density profile
using equation~(\ref{eq:exv}), it provides a direct link between
the density profile and interfacial thermodynamics, as measured by
the surface free energy. In fact, the overestimation of
$v_\textrm{ex}$ by the FMT at high packing fractions is dominated
by a significant underestimation, at high $\eta$ by the FMT, of
the second and third peaks of $\rho(z)$ (figure~\ref{fig:profile3}
and ~\ref{fig:profile4}). Because the pressure is a monotonously
increasing function of $\eta$, the overestimation of
$v_\textrm{ex}$ at high packing fractions by the FMT leads
directly, through equation (\ref{eq:dgdp}) to the observed
overestimation of $\gamma$ by the FMT.

\section{Discussion}
\label{sec:4}

The overall agreement between the benchmark simulation data for
the density profiles, the wall surface free energy and the excess
adsorption of hard sphere fluid at a planar hard wall and the
results obtained using accurate FMT DFT calculation is very good.
We have found, however, that systematic deviations develop at
sufficiently large packing fractions $\eta$. First, small
deviations develop at around $\eta\geqslant 0.3$. They become more
serious for $\eta\geqslant 0.4$ and show up in all quantities that
we have studied here, i.e., the density profiles, the surface free
energy and the excess adsorption. This was also observed in a
study of confined hard-sphere fluids \cite{Deb11}.

We find that for the dividing surface employed in our study, the
actual hard wall position, the DFT results of the surface free
energy $\gamma$ is less sensitive to the deviations in the density
profiles than the excess adsorption $\Gamma$. While the deviation
in $\gamma$ between  the DFT and the simulation results is  small,
even at high packing fractions, the DFT results for $\Gamma$ show
far larger relative deviations
 compared to the benchmark simulations. At the moment it is not clear
whether these deviations can be reduced by changing the form of the
excess free energy density $\Phi$~--- the functional form of the
empirical parametrization of $\gamma$ and $\Gamma$ \cite{Davidchack15}
could give a hint~--- or if they are due to the range of weight
functions employed by FMT.

\section*{Acknowledgements}
BBL acknowledges support from the National Science Foundation (NSF) under grant CHE-1465226.  This research used the ALICE High Performance Computing Facility at the University of Leicester.

\clearpage

\ukrainianpart

\title{Тверді сфери біля плоскої твердої стіни: моделювання та теорія функціоналу
густини}
\author{Р.Л. Давидчак\refaddr{label1}, Б.Б. Лейрд\refaddr{label2}, Р. Рот\refaddr{label3}}
\addresses{
\addr{label1} Математичний факультет, Університет Лестера, Лестер, LE1 7RH, Великобританія
\addr{label2} Хімічний факультет, Університет Канзасу, Лоуренс, Канзас 66045, США
\addr{label3} Інститут теоретичної фізики, Університет Тюбінґена, D-72074 Тюбінґен, Німеччина}

\makeukrtitle

\begin{abstract}
\tolerance=3000%
Тверді сфери є центральною і важливою моделлю системи відліку для однорідних і
неоднорідних плинних систем. У цій статті ми представляємо нові високоточні моделювання
методом молекулярної динаміки для плину твердих сфер біля плоскої твердої стінки.
Для цієї системи ми представляємо дані по профілю густини $\rho(z)$ для різних
об'ємних густин, поверхневої вільної енергії $\gamma$, надлишкової адсорбції $\Gamma$
і надлишкового об'єму $v_\text{ex}$, який тісно пов'язаний з $\Gamma$. Ми порівнюємо
всі величини з передбаченнями з оригінальних розрахунків теорії класичного функціоналу
густини в рамках теорії фундаментальної міри. Хоча ми отримуємо вцілому добре узгодження
між комп'ютерним моделюванням і теорією, значні відхилення появляються при достатньо високих об'ємних густинах.

\keywords неоднорідні плини, границі розділу тверде тіло-рідина, термодинаміка поверхні,
класична теорія функціоналу густини, моделювання методом молекулярної динаміки

\end{abstract}


\begin{thebibliography}{99}

\bibitem{Laird07} Laird B.B., Davidchack R.L., J. Phys. Chem., 2007, \textbf{111},
15935; \bibdoi{10.1063/1.1563248}.

\bibitem{Laird10} Laird B.B., Davidchack R.L., J. Chem. Phys., 2010, \textbf{132},
204101; \bibdoi{10.1063/1.3428383}.

\bibitem{Rosenfeld89}  Rosenfeld Y., Phys. Rev. Lett., 1989, \textbf{63}, 980; \bibdoi{10.1103/PhysRevLett.63.980}.

\bibitem{Roth10} Roth R., J. Phys.: Condens. Matter, 2010, \textbf{22}, 063102; \bibdoi{10.1088/0953-8984/22/6/063102}.

\bibitem{RapaportBook}  Rapaport D.C., {The Art of Molecular Dynamics Simulation}, 2nd Edn., Cambridge University Press, New York, 2004.

\bibitem{supplement} Supplementary data for this work can be found at  \url{http://hdl.handle.net/2381/33537}.

\bibitem{Laird12} Laird B.B.,  Hunter A., Davidchack R.L., Phys. Rev. E,  2012, \textbf{86}, 060602(R); \bibdoi{10.1103/PhysRevE.86.060602}.

\bibitem{Courtemanche92}  Courtemanche D.J.,  van Swol F., Phys. Rev. Lett., 1992,  \textbf{69} 2078; \bibdoi{10.1103/PhysRevLett.69.2078}.

\bibitem{Dijkstra04}  Dijkstra M., Phys. Rev. Lett., 2004, \textbf{93}, 108303;
\bibdoi{10.1103/PhysRevLett.93.108303}.

\bibitem{Auer03}  Auer S.,  Frenkel D., Phys. Rev. Lett., 2003, \textbf{91}, 015703;
\bibdoi{10.1103/PhysRevLett.91.015703}.

\bibitem{Evans79}  Evans R., Adv. Phys., 1979,  \textbf{28}, 143; \bibdoi{10.1080/00018737900101365}.

\bibitem{Bryk03}  Bryk P., Roth R.,  Mecke K.R.,  Dietrich S., Phys. Rev. E, 2003,  \textbf{68}, 031602; \bibdoi{10.1209/epl/i2004-10410-4}.

\bibitem{Roth02} Roth R.,  Evans R.,  Lang A.,  Kahl G., J. Phys.:
  Condens. Matter, 2002, \textbf{14}, 12063; \bibdoi{10.1088/0953-8984/14/46/313}.

\bibitem{Yu02}  Yu Y.-X.,  Wu J., J. Chem. Phys., 2002, \textbf{117}, 10156;  \bibdoi{10.1063/1.1520530}.

\bibitem{HansenGoos06b}  Hansen-Goos H., Roth R., J. Phys.: Condens. Matter, 2006,
  \textbf{18}, 8413; \bibdoi{10.1088/0953-8984/18/37/002}.


\bibitem{Mansoori71}  Mansoori G.A.,  Carnahan N.F.,  Starling K.E.,  Leland T.W. Jr.,
J. Chem. Phys., 1971, \textbf{54}, 1523; \bibdoi{10.1063/1.1675048}.

\bibitem{HansenGoos06a}  Hansen-Goos H., Roth R., J. Chem. Phys., 2006, \textbf{124},
  154506; \bibdoi{10.1063/1.2187491}.

\bibitem{Carnahan69}  Carnahan N.F.,  Starling K.E., J. Chem. Phys., 1969, \textbf{51},
  635; \bibdoi{10.1063/1.1672048}.

\bibitem{Boublik70}  Boubl\'ik T., J. Chem. Phys., 1970, \textbf{53}, 471; \bibdoi{10.1063/1.1673824}.

\bibitem{Leithall11}  Leithall G.,  Schmidt M., Phys. Rev. E, 2011, \textbf{83},
  021201; \bibdoi{10.1103/PhysRevE.83.021201}.

\bibitem{Korden12}  Korden S., Phys. Rev. E, 2012, \textbf{85}, 041150;  \bibdoi{10.1103/PhysRevE.85.041150}.

\bibitem{Marechal14}  Marechal M.,  Korden S.,  Mecke K., Phys. Rev. E, 2014,
  \textbf{90}, 042131; \bibdoi{10.1103/PhysRevE.90.042131}.

\bibitem{Davidchack15}  Davidchack R.L.,  Laird B.B., Roth R.,
  Mol. Phys., 2015,  \textbf{113}, 1091; \bibdoi{10.1080/00268976.2014.986240}.

\bibitem{Deb11}  Deb D.,  Winkler A.,  Yamani M.H.,  Oettel M.,
 Virnau P.,  Binder K., J. Chem. Phys., 2011, \textbf{134}, 214706;  \\ \bibdoi{10.1063/1.3593197}.

\end{thebibliography}
\end{document}